# Discrete Dynamics in Nature and Society

# Adaptive guaranteed-performance consensus control for multi-agent systems with an adjustable convergence speed

Haiying Ma[1], Xiao Jia[2], Ning Cai[3], Jianxiang Xi[2,*]

1. School of Economics, Northwest Minzu University, Lanzhou 730030
2. Rocket Force University of Engineering, Xi'an 710025
3. School of Automation, Beijing University of Posts and Telecommunications, Beijing 100876

E-mail: xijx07@mails.tsinghua.edu.cn

## Abstract

Adaptive guaranteed-performance consensus control problems for multi-agent systems are investigated, where the adjustable convergence speed is discussed. This paper firstly proposes a novel adaptive guaranteed-performance consensus protocol, where the communication weights can be adaptively regulated. By the state space decomposition method and the stability theory, sufficient conditions for guaranteed-performance consensus are obtained, as well as the guaranteed-performance cost. Moreover, since the convergence speed is usually adjusted by changing the algebraic connectivity in existing works, which increases the communication burden and the load of the controller, and the system topology is always given in practical applications, the lower bound of the convergence coefficient for multi-agent systems with the adaptive guaranteed-performance consensus protocol is deduced, which is linearly adjustable approximately by changing the adaptive control gain. Finally, simulation examples are introduced to demonstrate theoretical results.

## Introduction

In recent years, by the incentive effects of spacious applications, such as synchronization[1,2], formation control[3-5], cluster flight[6-8] and other fields[9-11], there is considerable attention in distributed cooperative control of multi-agent systems. As a significant topic in cooperative control, consensus, which means that all agents in a multi-agent system achieve an agreement on some factors by designing reasonable consensus control protocols, arouses extensive interest from investigators, and some meaningful relevant works have been developed in [12-15], where the consensus performance is not taken into account.

In some practical complex multi-agent systems, except for the requirements of achieving consensus, the consensus performance also need to be taken into consideration. There is a representative example in [16], and it is shown that when a particular task is carried out by multiple mobile vehicles, the distance performance is likely to be a critical factor due to restricted resource. Generally speaking, when consensus should be achieved in multi-agent systems and under the condition that certain cost functions included in constraints are identified as minimum or maximum, one can model these correlative issues as optimal or suboptimal consensus. It should be pointed out that researchers always use guaranteed-performance consensualization to realize optimal consensus control. In order to deal with the optimal consensus problem for first-order multi-agent systems with both continuous-time and discrete-time dynamics, LQR-based optimal algorithms were proposed and the optimal Laplacian matrices were deduced in [17]. Guan et al.[18] studied the guaranteed-performance consensus problem for second-order multi-agent systems, and a performance function was provided for evaluating the performance of each agent based on impulsive control methods. Furthermore, aiming at high-order multi-agent systems, some researchers also introduced effective guaranteed-performance consensus algorithms to solve existing problems in [19-22]. Although many innovative and significant results were given in [16-22], the convergence speed of multi-agent systems was not considered, which was one of the most important aspects for the system evaluation.

As is well known, the consensus speed was firstly referred by Olfati-Saber in [23] and [24]. Meanwhile, the relationship between the consensus speed and the algebraic connectivity of the system topology was presented, which was given the formal definition in [25]. It is no doubt that in view of the algebraic connectivity, many researchers have made much effort on improving the convergence speed by the transformation or reconstruction of topology structures shown in [26-30]. Ultra-fast consensus was achieved in small-world networks transformed from regular networks in [26]. An iterative algorithm employing semi-definite programming was applied in [27]. Moreover, based on hierarchical structures mentioned in [28] and [29], a multi-layer hierarchical structure algorithm was proposed in [30-32], which made it available to adjust the convergence speed. However, in practice, the system topology is always determined. In this case, optimizing consensus control protocols is a more useful way to change the algebraic connectivity with given topology structures. Yuan et al.[33] used partial information of second-order neighbors in different cases to accelerate distributed consensus through choosing suitable edges, and a similar approach was applied in [34]. In order to accelerate the convergence speed and reduce the communication cost among nodes with undirected topologies, Pseudo multi-hop relay algorithms were proposed in [35], which was also extended to high-order systems with directed topologies in [36]. In addition, predictive control methods were introduced to make multi-agent systems achieve fast consensus in [37] and [38]. It should be pointed out that the above interesting results did not consider the guaranteed-performance condition.

The current paper studies adaptive guaranteed-performance consensus control for multi-agent systems with an adjustable convergence speed. As far as I am concerned, the contribution of this paper is the following threefold: (i) A new adaptive guaranteed-performance consensus protocol for first-order multi-agent systems is proposed, and the communication weights among nodes in the system topology can be adaptively adjustable by state errors between each agent and its neighbors; (ii) The multi-agent system is decomposed into two subsystems which determine the consensus motion and the disagreement motion, respectively. Sufficient conditions for guaranteed-performance consensus are obtained, and the guaranteed-performance cost is determined at the same time; (iii) The convergence coefficient is defined for multi-agent systems under an adaptive consensus protocol, and the lower bound of the convergence coefficient is determined, which is related to the adaptive control gain and the minimum nonzero eigenvalue.

Compared with the existing relevant works on the consensus problem of multi-agent systems, this paper has following novelties. Firstly, a novel consensus control strategy is proposed to guarantee the consensus performance of multi-agent systems. Note that it corresponds with practical applications, which means that many systems in aviation and aerospace are constrained by limited energy. However, this method is not available in [12-15]. Secondly, the communication weights are regulated by both state errors and the adaptive control gain, while the communication weights cannot be regulated in [12-15]. Thirdly, in view of the given system topology and the huge communication burden, by changing the adaptive control gain rather than the algebraic connectivity related to the topology network, the convergence speed can be adjusted in this paper, but the adjustable convergence speed is not mentioned in [12-15].

The remainder of the paper is organized as follows. Section 2 states several important conclusions of graph theory and the problem description in which the adaptive guaranteed-performance consensus protocol is proposed. Section 3 shows sufficient conditions which can ensure the guaranteed-performance consensus of multi-agent systems and the convergence coefficient associated with the convergence speed. The theoretical results are illustrated via simulation examples in section 4. Finally, section 5 draws the conclusion.

*Notations*: $R^N$ is the real column vector space of dimension $N$, and $R^{N \times N}$ denotes the set of $N \times N$ dimensional real matrices. The transpose of a matrix $Q$ is denoted as $Q^T$. $I_M$ stands for the identity matrix with dimension $M$. $\mathbf{0}$ represents the zero column vector with appropriate dimension. Let $\mathbf{1}_M = [1, \cdots, 1]^T$ be the vector of $M$ ones with all components 1.

## Materials and Methods

### Preliminaries and problem description

This section mainly introduces some basic concepts of graph theory and presents the problem description.

#### A. Algebraic graph theory

A connected undirected graph $G = (V, E, W)$ can be used to describe the information interchange among agents in a multi-agent system, where $V = \{v_1, v_2, \cdots, v_N\}$, $E = V \times V$ and $W = [w_{ij}] \in R^{N \times N}$ stand for a set of vertices, a set of edges and the adjacency matrix related to the graph $G$, respectively. $e_{ij} = (v_i, v_j) \in E$ represents an edge in $G$, where the information exchange exists between node $v_i$ and node $v_j$, and node $v_i$ is called a neighbor of node $v_j$. Note that we can describe all neighbors of node $v_j$ as an index set $N_j = \{i \mid (v_i, v_j) \in E\}$. For the undirected graph $G$, $(v_i, v_j) \Leftrightarrow (v_j, v_i)$. The adjacency element $w_{ij}$ corresponds to the interaction strength between node $v_i$ and node $v_j$, where $w_{ij} = 1$ shows $e_{ij} = (v_i, v_j) \in E$ and $w_{ij} = 0$ shows $e_{ij} = (v_i, v_j) \notin E$. It is assumed that there is no self edge for each node; that is, $w_{ii} = 0$. Defined by $L = [l_{ij}] \in R^{N \times N}$ the Laplacian matrix of $G$, where $l_{ii} = \sum_{j=1, j \neq i}^{N} w_{ij}$ and $l_{ij} = -w_{ij} (i \neq j)$. According to the definition of connected undirected graphs, one eigenvalue of $L$ is zero, and the remaining eigenvalues are positive.

#### B. Problem description

Consider a complex multi-agent system with $M$ nodes with each node modeled as
$$\dot{x}_k(t) = u_k(t) \quad (k = 1, 2, \cdots, M), \tag{1}$$
where $x_k(t) \in \mathbf{R}$ is the state of the $k$ th node to be coordinated and $u_k(t) \in \mathbf{R}$ is the control input of the $k$ th node to be designed according to the information of its neighboring nodes. The system topology among nodes is modeled as a connected undirected graph.

The following adaptive guaranteed-performance control protocol is proposed:

$$\begin{cases} u_k(t) = \alpha \sum_{j \in N_k} \beta_{kj} w_{kj}(t)\left(x_j(t) - x_k(t)\right), \\ \dot{w}_{kj}(t) = \alpha \left(x_j(t) - x_k(t)\right)^2, \\ J_r = \dfrac{1}{M} \sum_{k=1}^{M} \sum_{j=1}^{M} \int_0^{+\infty} \zeta \left(x_j(t) - x_k(t)\right)^2 \mathrm{d}t, \end{cases} \quad (2)$$

where $\alpha$ is the adaptive control gain, $\zeta$ represents the performance coefficient with $\zeta > 0$, $N_k$ denotes the neighbor set of node $k$, $\beta_{kj}$ stands for the connection state of the edge between node $j$ and node $k$, which equals to one if node $j$ is connected to node $k$ and identically equals to zero otherwise, and $w_{kj}(t)$ and $\dot{w}_{kj}(t)$ are the adaptively regulated weight of the edge from node $j$ to node $k$ and its time derivative, respectively.

It can be found that if $\alpha > 0$, $w_{kj}(t)$ is non-decreasing and time-varying and its variation is closely associated with the state errors between two agents. Furthermore, the growth rate of $w_{kj}(t)$ is faster when the state errors are greater, which means that the greater effect is given to adjust the state errors between them. Especially, $w_{kj}(t)$ is not changed when the states of two agents are equal. Without loss of generality, it is assumed that the initial value of weight $w_{kj}(0)$ is 1 and the upper bound of $w_{kj}(t)$ is $w_{kjm}$.

In consideration of the undirected topology $G$, one can easily see that nodes in the topology correspond to agents of multi-agent systems. In this case, let $\boldsymbol{x}(t) = [x_1(t), x_2(t), \cdots, x_M(t)]^T$, then a compact form of system (1) with control protocol (2) is produced as follows

$$\dot{\boldsymbol{x}}(t) = -\alpha L_{w(t)} \boldsymbol{x}(t), \quad (3)$$

where $L_{w(t)}$ stands for the Laplacian matrix of $G$.

The definition of the adaptive guaranteed-performance consensus for multi-agent system (1) is described as follows.

***Definition 1.*** Combined with control protocol (2), multi-agent system (1) is said to achieve adaptive guaranteed-performance consensus if there exists $\alpha$ such that $\lim_{t \to \infty}(x_k(t) - x_j(t)) = 0$ ($k, j = 1, 2, \cdots, M$), and $J_r \leq J_r^*$ in view of any bounded initial states $x_k(0)$ ($k = 1, 2, \cdots, M$), where $J_r^*$ is known as the guaranteed-performance cost.

***Remark 1.*** The aim of this paper is to obtain a suitable adaptive control gain and the guaranteed-performance cost, so that multi-agent system (1) can achieve adaptive guaranteed-performance consensus. Moreover, there are two main features in adaptive guaranteed-performance control protocol (2). Firstly, the guaranteed-performance function $J_r$ associated with the performance coefficient and state errors is proposed. In this case, the consensus performance of multi-agent systems is guaranteed, which is more meaningful compared with [12-15]. Secondly, the adaptive control gain is designed, and it can effectively regulate the time-varying communication weights, which plays an important role in the consensus analysis for multi-agent systems. However, this factor is not referred in [12-15].

**Main results**

In the following theoretical analysis, sufficient conditions for multi-agent system (1) to achieve adaptive guaranteed-performance consensus are obtained by designing the adaptive control gain $\alpha$, then the guaranteed-performance cost $J_r^*$ is determined. Moreover, the consensus convergence speed is discussed, and the lower bound of the convergence coefficient is presented to indicate the convergence speed. We prove that it is an effective way to linearly regulate the consensus convergence speed by changing $\alpha$.

**C. adaptive guaranteed-performance consensus analysis**

In the first place, the state space decomposition approach is proposed to transform the dynamics of multi-agent system (1). Denoted by $0 = \lambda_1 < \lambda_2 \leq \cdots \leq \lambda_M$ the eigenvalues of $L_{w(0)}$. Then according to matrix theory, one can find that there exists an orthogonal matrix $Q = [\mathbf{1}_M/\sqrt{M} \quad \tilde{Q}]$ with $\mathbf{1}_M = [1,1,\cdots,1]^T \in \mathbf{R}^M$ satisfying

$$Q^T L_{w(0)} Q = \begin{bmatrix} 0 & \mathbf{0}^T \\ \mathbf{0} & \Omega_{(0)} \end{bmatrix}, \tag{4}$$

where $\Omega_{(0)} = \tilde{Q}^T L_{w(0)} \tilde{Q} = \mathbf{diag}\{\lambda_2, \lambda_3, \cdots, \lambda_M\}$. Let

$$\tilde{\mathbf{x}}(t) = Q^T \mathbf{x}(t) = [\tilde{x}_1^T(t), \boldsymbol{\mu}^T(t)]^T, \tag{5}$$

where $\boldsymbol{\mu}(t) = [\tilde{x}_2^T(t), \tilde{x}_3^T(t), \cdots, \tilde{x}_M^T(t)]^T$, then one can get that multi-agent system (3) can be rewritten as

$$\dot{\tilde{x}}_1(t) = 0, \tag{6}$$

$$\dot{\boldsymbol{\mu}}(t) = -\alpha \tilde{Q}^T L_{w(t)} \tilde{Q} \boldsymbol{\mu}(t). \tag{7}$$

Let $\boldsymbol{e}_i (i = 1,2,\cdots,M)$ denote $M$-dimensional unit vectors with the $i$th element 1 and 0 elsewhere. Define

$$\mathbf{x}_c(t) := Q\boldsymbol{e}_1 \tilde{x}_1(t) = \frac{1}{\sqrt{M}} \tilde{x}_1(t), \tag{8}$$

$$\mathbf{x}_{\bar{c}}(t) := \sum_{i=2}^{M} Q\boldsymbol{e}_i \otimes \tilde{x}_i(t). \tag{9}$$

Due to

$$\sum_{i=2}^{M} \boldsymbol{e}_i \otimes \tilde{x}_i(t) = [\mathbf{0}, \boldsymbol{\mu}^T(t)]^T, \tag{10}$$

one can obtain from (8) that

$$\mathbf{x}_c(t) = Q[\tilde{x}_1^T(t), \mathbf{0}]^T. \tag{11}$$

Similar to the above analysis, it follows from (9) that

$$\mathbf{x}_{\bar{c}}(t) = Q[\mathbf{0}, \boldsymbol{\mu}^T(t)]^T. \tag{12}$$

According to the definition of orthogonal matrices, one can see that $Q$ is nonsingular. Thus it holds that $\mathbf{x}_c(t)$ and $\mathbf{x}_{\bar{c}}(t)$ are linearly independent with each other by (11) and (12). Note that $\tilde{\mathbf{x}}(t) = Q^T \mathbf{x}(t) = [\tilde{x}_1^T(t), \boldsymbol{\mu}^T(t)]^T$, it can be shown that

$$\mathbf{x}(t) = \mathbf{x}_c(t) + \mathbf{x}_{\bar{c}}(t). \tag{13}$$

Due to the form of $\mathbf{x}_c(t)$ in (8), one can show that multi-agent system (1) under control protocol (2) can be adaptively guaranteed-performance consensualizable if and only if $\lim_{t \to \infty} \boldsymbol{\mu}(t) = 0$; in other words, subsystems (6) and (7) are presented respectively to indicate the consensus motion and relative state motion of multi-agent system (1).

In the following theorem, sufficient conditions for adaptive guaranteed-performance consensus are obtained, which means that distributed guaranteed-performance consensus design for multi-agent system (1) is derived.

***Theorem 1.*** Multi-agent (1) with control protocol (2) achieves adaptive guaranteed-performance consensus if $\alpha \geq 2\zeta$. In this case, the guaranteed-performance cost satisfies that

$$J_r^* = \boldsymbol{x}^T(0)\left(\boldsymbol{I}_M - \frac{1}{M}\boldsymbol{1}_M\boldsymbol{1}_M^T\right)\boldsymbol{x}(0) + \alpha\int_0^{+\infty}\boldsymbol{x}^T(t)\left(\boldsymbol{I}_M - \frac{1}{M}\boldsymbol{1}_M\boldsymbol{1}_M^T\right)\boldsymbol{x}(t)\mathrm{d}t.$$

***Proof.*** To begin with, we prove $\lim_{t\to\infty}\boldsymbol{\mu}(t) = \boldsymbol{0}$ under $\alpha \geq 2\zeta > 0$. Consider a Lyapunov function candidate as follows.

$$V(t) = \boldsymbol{\mu}^T(t)\boldsymbol{\mu}(t) + \sum_{k=1}^{M}\sum_{j=1}^{M}\frac{(w_{kj}(t) - w_{kj}(0))^2}{2} + \frac{1}{2M}\sum_{k=1}^{M}\sum_{j=1}^{M}(w_{kjm} - w_{kj}(t)). \tag{14}$$

Because of $w_{kjm} \geq w_{kj}(t)$, one can get that $V(t) \geq 0$ in spite of the value of $\alpha$. Furthermore, taking the time derivative of $V(t)$ along the trajectory of subsystem (7), one has

$$\dot{V}(t) = (-\alpha\boldsymbol{\mu}^T(t)\tilde{Q}^T L_{w(t)}^T\tilde{Q}\boldsymbol{\mu}(t) - \boldsymbol{\mu}^T(t)\alpha\tilde{Q}^T L_{w(t)}\tilde{Q}\boldsymbol{\mu}(t)) + \sum_{k=1}^{M}\sum_{j=1}^{M}(w_{kj}(t) - w_{kj}(0))\dot{w}_{kj}(t)$$

$$- \frac{1}{2M}\sum_{k=1}^{M}\sum_{j=1}^{M}\dot{w}_{kj}(t). \tag{15}$$

Due to the assumption of the undirected connected graph $G$, it can be seen that $L_{w(t)}^T = L_{w(t)}$. Then since equation $QQ^T = I_M$ holds, one can easily obtain $\tilde{Q}\tilde{Q}^T = L_M$, where $L_M$ stands for the Laplacian matrix of a complete graph with the weights of all edges $1/M$. Moreover, one has

$$\frac{1}{2}\sum_{k=1}^{M}\sum_{j=1}^{M}w_{kj}(t)(x_j(t) - x_k(t))^2 = \boldsymbol{x}(t)^T L_{w(t)}\boldsymbol{x}(t), \tag{16}$$

Then by (2) and (16), it can be seen that

$$\sum_{k=1}^{M}\sum_{j=1}^{M}(w_{kj}(t) - w_{kj}(0))\dot{w}_{kj}(t) = 2\alpha\boldsymbol{x}^T(t)L_{w(t)}\boldsymbol{x}(t) - 2\alpha\boldsymbol{x}^T(t)L_{w(0)}\boldsymbol{x}(t). \tag{17}$$

In addition, it can be also obtained that

$$\frac{1}{2M}\sum_{k=1}^{M}\sum_{j=1}^{M}\dot{w}_{kj}(t) = \alpha\boldsymbol{x}^T(t)L_M\boldsymbol{x}(t). \tag{18}$$

Substituting (17) and (18) into (15), it can be deduced that

$$\dot{V}(t) = -2\alpha\boldsymbol{\mu}^T(t)\tilde{Q}^T L_{w(t)}\tilde{Q}\boldsymbol{\mu}(t) + 2\alpha\boldsymbol{x}^T(t)L_{w(t)}\boldsymbol{x}(t) - 2\alpha\boldsymbol{x}^T(t)L_{w(0)}\boldsymbol{x}(t) - \alpha\boldsymbol{x}^T(t)L_M\boldsymbol{x}(t)$$

$$= -2\alpha\boldsymbol{\mu}^T(t)\tilde{Q}^T L_{w(0)}\tilde{Q}\boldsymbol{\mu}(t) - \alpha\boldsymbol{\mu}^T(t)\tilde{Q}^T L_M\tilde{Q}\boldsymbol{\mu}(t) \tag{19}$$

$$= -\alpha\sum_{k=2}^{M}\tilde{x}_i^T(t)(2\lambda_k + 1)\tilde{x}_k(t).$$

Moreover, it should be pointed that $\lambda_k > 0\ (k = 2,3,\cdots,M)$, then if $\alpha > 0$, the following inequality holds

$$\dot{V}(t) \leq -\alpha(2\lambda_2 + 1)\sum_{k=2}^{M}\tilde{x}_k^T(t)\tilde{x}_k(t) = -\alpha(2\lambda_2 + 1)\|\boldsymbol{\mu}(t)\|^2. \tag{20}$$

Accordingly, $\boldsymbol{\mu}(t)$ converges to $\boldsymbol{0}$; that is, multi-agent system (1) under control protocol (2) can be adaptively consensusalizable.

In the following discussion, the guaranteed-performance cost is determined. Firstly, One can obtain from (15) that

$$\sum_{k=1}^{M}\sum_{j=1}^{M}(x_j(t)-x_k(t))^T \zeta (x_j(t)-x_k(t)) = 2M\zeta x^T(t)L_M x(t) = 2M\zeta \mu^T(t)\tilde{Q}^T L_M \tilde{Q}\mu(t)$$
$$= 2M\zeta \sum_{k=2}^{M}\tilde{x}_k^T(t)\tilde{x}_k(t). \quad (21)$$

Let $H > 0$. Then define

$$J_r^H \triangleq \frac{1}{M}\sum_{k=1}^{M}\sum_{j=1}^{M}\int_0^H (x_j(t)-x_k(t))^T \zeta (x_j(t)-x_k(t))\mathrm{d}t, \quad (22)$$

Substituting (21) into (22) leads to

$$J_r^H = 2\zeta \int_0^H \sum_{k=2}^{M}\tilde{x}_k^T(t)\tilde{x}_k(t)\mathrm{d}t. \quad (23)$$

Furthermore, it can be shown that

$$\int_0^H \dot{V}(t)\mathrm{d}t - V(H) + V(0) = 0. \quad (24)$$

Since $\lim_{t\to\infty}(w_{kj}(t)-w_{kjm}) = 0$, one can easily get that

$$\lim_{H\to\infty}\sum_{k=1}^{M}\sum_{j=1,k\neq j}^{M}(w_{kj}(H)-w_{kjm}) = 0. \quad (25)$$

Note that $\dot{V}(t) = -\alpha \sum_{k=2}^{M}\tilde{x}_k^T(t)(2\lambda_k+1)\tilde{x}_k(t) \leq -\alpha \sum_{k=2}^{M}\tilde{x}_k^T(t)\tilde{x}_k(t)$. Then if $\alpha \geq 2\zeta$, it follows from (23),(24) and (25) that

$$\lim_{H\to\infty}J_r^H = \lim_{H\to\infty}2\zeta\int_0^H \sum_{k=2}^{M}\tilde{x}_k^T(t)\tilde{x}_k(t)\mathrm{d}t \leq \mu^T(0)\mu(0) + \frac{1}{2M}\sum_{k=1}^{M}\sum_{j=1,k\neq j}^{M}(w_{kjm}-w_{kj}(0)). \quad (26)$$

Due to $\mu(t) = \tilde{Q}^T x(t)$ and $\tilde{Q}\tilde{Q}^T = L_M$, one gets

$$\mu^T(0)\mu(0) = x^T(0)\left(I_M - \frac{1}{M}\mathbf{1}_M\mathbf{1}_M^T\right)x(0). \quad (27)$$

Since $\lim_{t\to\infty}(w_{kj}(t)-w_{kjm}) = 0$, one has

$$\sum_{k=1}^{M}\sum_{j=1,k\neq j}^{M}(w_{kjm}-w_{kj}(0)) = \sum_{k=1}^{M}\sum_{j=1,k\neq j}^{M}\int_0^{+\infty}\dot{w}_{kj}(t)\mathrm{d}t = 2M\alpha\int_0^{+\infty}x^T(t)\left(I_M - \frac{1}{M}\mathbf{1}_M\mathbf{1}_M^T\right)x(t)\mathrm{d}t. \quad (28)$$

From (26) to (28), the proof of Theorem 1 is completed.□

*Remark 2.* It can be seen from Theorem 1 that the relationship between the adaptive control gain and the performance coefficient is obtained, and the guaranteed-performance cost related to the adaptive control gain is determined, which means that the guaranteed-performance cost can be regulated by changing the adaptive control gain. Then by the adaptive control strategy, the impacts of time-varying communication weights can be reduced in the adaptive guaranteed-performance consensus analysis. However, the performance function is not mentioned in [12-15], let alone the guaranteed-performance cost. In this case, the consensus performance of multi-agent systems cannot be guaranteed, which has less significance in practical applications. Furthermore, although time-varying communication weights is considered in [13] and [14], there is no effective method to overcome its impacts; that is, the consensus analysis is more complicated affected by time-varying communication weights.

### D. Convergence speed analysis

It should be noted that the Laplacian matrix $L$ related to the undirected graph $G$ is a semi-definite matrix. Then being the minimum nonzero eigenvalue of $L$, $\lambda_2$ is also said to be the algebraic connectivity satisfying

$$\lambda_2 = \min_{x \neq 0, 1_M^T x = 0} \frac{\boldsymbol{x}^T(t) L \boldsymbol{x}(t)}{\|\boldsymbol{x}(t)\|^2}. \tag{29}$$

According to [23], one can see that in case there is no gain matrix in the consensus control protocol, the convergence speed is associated with the minimum nonzero eigenvalue $\lambda_2$. In other words, the larger the value of $\lambda_2$ is, the faster the convergence speed is.

For the sake of verifying the assumption that the proposed method can effectively regulate the convergence speed of multi-agent system (1), we give the definition of the convergence coefficient as follows.

***Definition 2.*** For multi-agent system (1), the convergence coefficient $\rho(t)$ can be defined as

$$\rho(t) = -\frac{\dot{V}(t)}{\tilde{V}(t)}. \tag{30}$$

where $V(t)$ is the Lyapunov function, $\dot{V}(t)$ is the time derivative of $V(t)$, and $\tilde{V}(t) = \boldsymbol{\mu}^T(t)\boldsymbol{\mu}(t)$. Then the lower bound of $\rho(t)$ is shown as $\rho_{\min}$ which means the minimum convergence speed of multi-agent system (1).

As a matter of fact, equation (28) originates from Definition 2. Then with regard to multi-agent system (1) under the standard control protocol in [23], where $V(t) = \boldsymbol{x}^T(t)\boldsymbol{x}(t)$ and $\dot{V}(t) = -2\boldsymbol{x}^T(t)L\boldsymbol{x}(t)$, the convergence coefficient can be expressed as

$$\rho(t) = -\frac{\dot{V}(t)}{\tilde{V}(t)} \geq 2\lambda_2. \tag{31}$$

It can be seen that $\rho_{\min} = 2\lambda_2$; that is, the lower bound of $\rho(t)$ is directly associated with the algebraic connectivity. Since there exists the relationship between $\rho_{\min}$ and the algebraic connectivity, it is rational and convenient to use $\rho_{\min}$ to describe the convergence speed to some extent.

In the following, we determine the convergence coefficient of multi-agent system (1) under control protocol (2) and compare it with the convergence coefficient under the standard consensus protocol. In order to ensure the effectiveness of this comparison, the control gain $\alpha$ is also considered in the standard consensus protocol as a reference, which means that $u_k(t) = \alpha \sum_{j \in N_k} w_{kj}(0)(x_j(t) - x_k(t))$. In this case, one can directly obtain that $\rho_{\min,1} = 2\alpha\lambda_2$.

Then substituting (17) into (19), the convergence coefficient of multi-agent system (1) under control protocol (2) can be described as

$$\rho(x) = -\frac{\dot{V}(x)}{\tilde{V}(x)} \geq \alpha(2\lambda_2 + 1). \tag{32}$$

Therefore, the following theorem can be obtained.

***Theorem 2.*** The lower bound of the convergence coefficient of multi-agent (1) under control protocol (2) is $\rho_{\min,2} = \alpha(2\lambda_2 + 1)$.

***Remark 3.*** As the improvement of the convergence speed can save working time of multi-agent systems in practice to some degree, it is significant to investigate how to improve the convergence speed when multi-agent systems achieve consensus. Thus, the adjustable convergence speed is deduced, and the lower bound of the convergence coefficient under control protocol (2) is derived in this paper, which means that one can regulate the convergence speed by changing the adaptive control gain or the algebraic connectivity. However, in [12-15], the convergence speed is not considered; that is, one cannot regulate the consensus convergence according to the requirements in different environment.

**Remark 4.** By comparison, one can see that although changing the value of $\alpha$ can regulate the lower bound $\rho_{\min 1}$, the minimum nonzero eigenvalue of $L$ is usually small, and cannot be determined precisely. Thus, it is also difficult to choose proper control gain to adjust $\rho_{\min 1}$. On the contrary, there exists an extra item $\alpha$ in equation $\rho_{\min,2} = \alpha(2\lambda_2 + 1)$, which means that $\alpha$ can play a more important role in achieving the adjustable convergence speed. In other words, with the increase of the value of $\alpha$, $\rho_{\min,2}$ is increased obviously, and is less affected by the minimum nonzero eigenvalue of $L$. In this case, the convergence speed can be promoted more quickly. Furthermore, $\rho_{\min,2}$ is linearly associated with $\alpha$ on account of fixed $\lambda_2$. As a result, the convergence speed of multi-agent system (1) can be linearly regulated approximately. However, although many approaches have been used to adjust the convergence speed in [33-38], all of them depend on the minimum nonzero eigenvalue of $L$ associated with the system topology.

## Results and Discussion

In this section, a simulation example is given to demonstrate the theoretical results shown in the previous analysis.

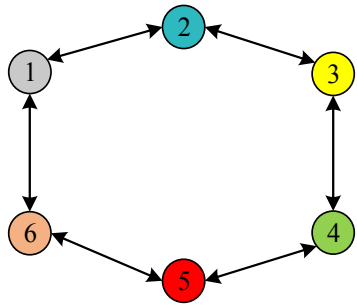

Fig. 1 system topology

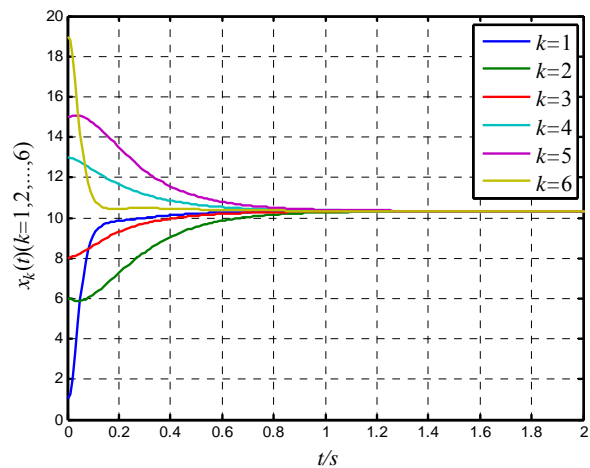

Fig. 2: State trajectories under control protocol (2)

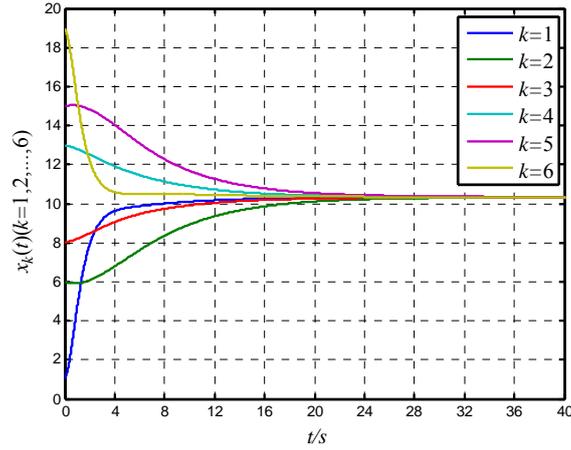

Fig. 3: State trajectories under the standard consensus protocol

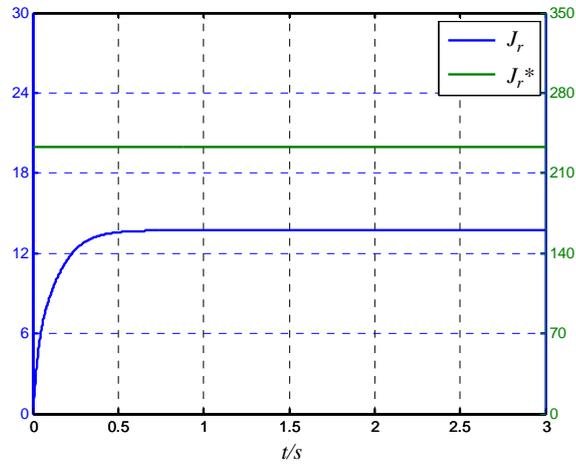

Fig. 4: Guaranteed-performance function of multi-agent system (1)

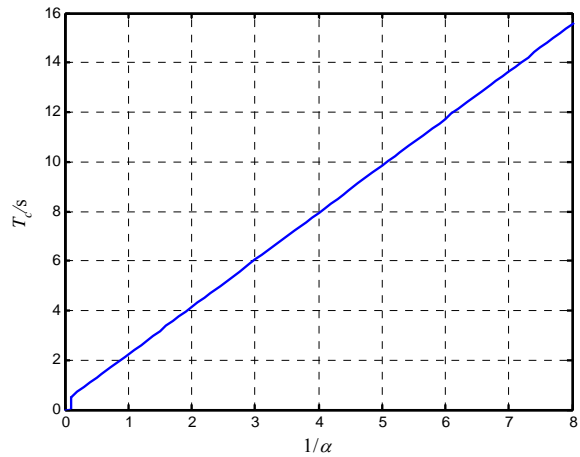

Fig. 5: The relationship between convergence time and reciprocal of adaptive control gain

Consider a first-order multi-agent system composed of six agents, where the system topology is shown in Fig. 1. The initial state of agents is $x(0) = [1, 6, 8, 13, 15, 19]^T$. Fig. 2 depicts the state trajectories of multi-agent (1) under control protocol (2), one can see that the states of all agents reach a common value, and compared with the state trajectories of multi-agent (1) under the standard consensus protocol in Fig. 3, the convergence time of the former is much less; that is, the convergence speed is faster. Fig. 4 depicts the guaranteed-performance function $J_r^H$ with $H=3$, and $J_r^H$ converges to a finite value with $J_r^H < J_r^*$. Fig. 5 shows the relationship between the convergence time and $1/\alpha$. One can obtain that the convergence time can be positively correlative approximately with $1/\alpha$, provided that $\alpha$ is small enough.

## Conclusions

A new adaptive guaranteed-performance consensus scheme for multi-agent systems with an adjustable convergence speed was proposed in this paper. The adaptive guaranteed-performance consensus protocol was presented by adjusting the communication weights among agents in the system topology. Sufficient conditions for adaptive guaranteed-performance consensus was obtained and the guaranteed-performance cost was deduced. Then in order to indicate the convergence speed of multi-agent systems, the convergence coefficient was defined, and it was proved that the convergence speed can be approximately linear adjustable by changing the adaptive control gain. Furthermore, another interesting work in the following years is to consider the external disturbance for the guaranteed-performance consensus of multi-agent systems.

## Conflicts of Interest

The authors declare that there is no conflict of interest regarding the publication of this paper.

## Funding Statement

This work is supported by National Natural Science Foundation under grant #61867005.

## Acknowledgments

I would like to express my faithful gratitude to all those who helped me during the writing of this paper. I gratefully acknowledge the help of my partners, Mr. Cai Ning and Mr. Xi Jianxiang, who have offered me patient instruction in the academic studies.

## Supplementary Materials

All the data in the simulation is stored in an Excel file which can be accessed from the https://pan.baidu.com/s/10Yj9jqNWWf4AR4c0GnWr-Q following link.